\documentclass[conference]{IEEEtran}
\IEEEoverridecommandlockouts
\usepackage{amsmath,amsfonts,amssymb,amsthm}
\usepackage{algorithm,algorithmic}
\usepackage{array,booktabs,multirow}
\usepackage{graphicx}
\usepackage{textcomp}
\usepackage[table]{xcolor}
\usepackage{xspace}
\usepackage{stfloats}
\usepackage{balance}
\usepackage{url}
\usepackage{verbatim}
\usepackage{cite}

\usepackage[caption=false,font=tiny]{subfig}

\newtheorem{myTheo}{Theorem}
\newtheorem{myAssu}{Assumption}
\newtheorem{myLem}{Lemma}

\def\BibTeX{{\rm B\kern-.05em{\sc i\kern-.025em b}\kern-.08em
    T\kern-.1667em\lower.7ex\hbox{E}\kern-.125emX}}
\begin{document}

\title{TSFLora: Token-Compressed Split Fine-Tuning for Wireless Edge Networks}

\author{
        \IEEEauthorblockN{Xianke Qiang\IEEEauthorrefmark{1}, 
        Zheng Chang\IEEEauthorrefmark{1},
        Li Wang\IEEEauthorrefmark{3},
        Ying-Chang Liang\IEEEauthorrefmark{2} }		 
    
    \IEEEauthorblockA{\IEEEauthorrefmark{1}School of Computer Science and Engineering, UESTC, 611731 Chengdu, China}
    
    \IEEEauthorblockA{\IEEEauthorrefmark{2}Center for Intelligent Networking and Communications, UESTC, 611731 Chengdu, China.}

     \IEEEauthorblockA{\IEEEauthorrefmark{3}School of Computer Science, Beijing University of Posts and Telecommunications, Beijing, 100876, China.}
}
\maketitle

\begin{abstract}
Adapting large AI models (LAMs) to personalized edge data is challenging because wireless devices have limited memory, computation, and uplink capacity. Federated fine-tuning preserves data privacy but still requires each device to host the full model, while split learning reduces device memory at the cost of heavy activation transmission. This paper proposes TSFLora, a token-compressed split fine-tuning framework for communication-efficient LAM adaptation at the edge. TSFLora combines attention-guided token selection, token merging, low-bit activation quantization, and LoRA-based adaptation within a split federated training pipeline. The key idea is to compress the intermediate token sequence before transmission so that the system reduces both uplink traffic and server-side processing without changing the frozen backbone. Experiments on ViT models over CIFAR-10, CIFAR-100, and TinyImageNet show that TSFLora achieves up to \textbf{6.8$\times$} communication reduction and \textbf{41\%} memory saving while maintaining competitive accuracy.
\end{abstract}


\section{Introduction}\label{section1}
    Recent large AI models (LAMs) have shown strong transferability across tasks and domains~\cite{10579546}, which makes them attractive for edge intelligence in privacy-sensitive applications such as healthcare~\cite{yu2024fincon} and finance~\cite{thirunavukarasu2023large}. At the same time, edge devices continuously generate personalized data that are valuable for local adaptation~\cite{lin2024splitlora}. A practical deployment therefore requires fine-tuning LAMs without centralizing data, while also respecting the memory, computation, and bandwidth limits of edge hardware.\par
    Federated learning (FL) addresses the privacy requirement by keeping data local and exchanging model updates~\cite{mcmahan2017communication}. However, direct FL remains difficult for LAM fine-tuning because each device must host the full backbone and store activations for backpropagation. Parameter-efficient fine-tuning (PEFT) methods, such as adapters, prompt tuning, and LoRA~\cite{houlsby2019parameter, li2023prompt, hu2021lora}, reduce the number of trainable parameters, but they do not remove the need to keep the backbone in memory. For many edge platforms, this requirement is still prohibitive, as illustrated in Table~\ref{tab:introduction}.\par
    \begin{figure}[t]
        \centering
        \includegraphics[width=0.9\linewidth]{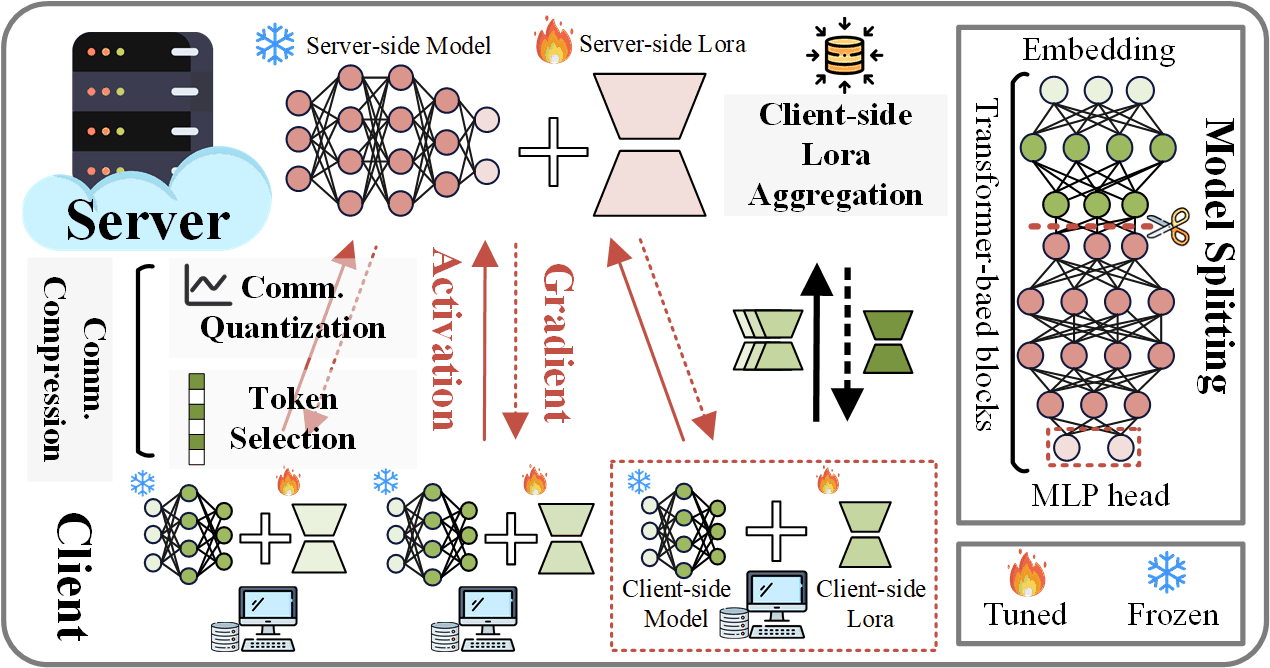}
        \caption{The overview of the TSFLora system.}
        \label{fig:framework}
    \end{figure}    
    Split Federated Learning (SFL) alleviates the device-side memory bottleneck by partitioning the model between devices and the server~\cite{thapa2022splitfed,10839234}. Each device executes only the front layers and uploads intermediate activations, while the server finishes the remaining forward and backward passes. This architecture lowers device memory usage, but it introduces two new system bottlenecks.\par

    First, uplink communication in SFL can be dominated by intermediate activations. In transformer models, these activations are token sequences whose size grows with the token number and embedding dimension. For ViT-B/16, the uplink traffic exceeds 233~MB per round in our setting (Table~\ref{tab:introduction}), even larger than the LoRA update payload in conventional FL. Therefore, reducing the size of transmitted activations is critical for improving communication efficiency.\par

    Second, server-side processing becomes a scalability bottleneck as more clients participate. The server must process uploaded activations through the remaining transformer blocks, and the cost increases with the sequence length. Therefore, reducing the transmitted token sequence can also lower server-side computation and memory overhead.\par

    Motivated by these challenges, we propose \textbf{TSFLora}, a token-compressed split fine-tuning framework for SFLv2\cite{thapa2022splitfed}. Token selection shortens the transmitted sequence and reduces server computation, quantization compresses each transmitted token, and LoRA keeps the number of trainable parameters small on both sides of the split model. Together, these components provide a practical way to balance accuracy and system efficiency under edge resource constraints. Our main contributions are summarized as follows:

    \begin{itemize}
        \item We propose \textbf{TSFLora}, a token-compressed split fine-tuning framework for communication-efficient adaptation of LAMs in wireless edge networks.
        \item We develop a two-stage activation compression scheme that combines attention-guided token selection and low-bit quantization to reduce uplink communication and server-side computation.
        \item We analyze the convergence of token-compressed split fine-tuning by characterizing the gradient perturbations caused by token selection and quantization.
        \item Experiments on ViT models with CIFAR-10, CIFAR-100, and TinyImageNet show that TSFLora achieves up to \textbf{6.8$\times$} communication reduction and \textbf{41\%} memory savings with little accuracy loss. The code is available at: \url{https://github.com/XiankeQiang/TSFLora}.
    \end{itemize}
    The rest of this paper is organized as follows. Section~\ref{section2} presents the system architecture and workflow. Section~\ref{section3} introduces the token compression mechanism at the token and bit levels. Section~\ref{section4} provides the non-convex convergence analysis. Section~\ref{section5} presents the problem formulation and design insights. Section~\ref{section6} describes the experimental setup, baselines, and results. Finally, Section~\ref{section7} concludes this paper.
    \begin{table}[t]
        \caption{Resource Gap Between Edge Devices and Large AI Models with Device-side Overhead}
        \label{tab:introduction}
        \centering
        \scriptsize
        \renewcommand{\arraystretch}{1.15}
        \setlength{\tabcolsep}{3pt}
        \resizebox{\linewidth}{!}{%
            \begin{tabular}{lc|llc|lcc}
                \toprule
                \multicolumn{2}{c|}{\textbf{Edge Device Memory}} & 
                \multicolumn{3}{c|}{\textbf{Model Demand}} & 
                \multicolumn{3}{c}{\textbf{Device-side LoRA Fine-tuning Overhead}} \\
                \textbf{Device} & \textbf{Memory} & \textbf{Model} & \textbf{\#Params} & \textbf{PEFT Memory} & \textbf{Paradigm} & \textbf{Comm.$^{1}$} & \textbf{Mem.$^{2}$} \\
                \midrule
                Raspberry Pi 5B    & 2/4/8 GB  & LLaMA-7B   & 7B   & 24 GB  & CL             & 229.2 MB   & 0 GB \\
                NVIDIA Jetson Nano & 4 GB      & LLaMA-65B  & 65B  & 260 GB & FL (ViT-B)     & 15.8 MB/R  & 4 GB \\
                NVIDIA Jetson TX2  & 8 GB      & ViT-B/32   & 86M  & 4 GB   & SFL (ViT-B/32) & 61.3 MB/R  & 1.1 GB \\
                NVIDIA Xavier NX   & 4/8/16 GB & ViT-L/32   & 330M & 8 GB   & SFL (ViT-B/16) & 233.5 MB/R & 2.3 GB \\
                \bottomrule
            \end{tabular}%
        }
    \end{table}

    
    \footnotetext[1]{The communication overhead of different schemes is compared under FP32 representation. For a $224 \times 224 \times 3$ image, the raw input size is $224 \times 224 \times 3 \times 4 = 602{,}112$ bytes ($0.602$ MB) per image. After the ViT-B/32 embedding layer, each image is represented by $50$ tokens, each with dimension $768$, resulting in $50 \times 768 \times 4 = 153{,}600$ bytes ($0.154$ MB) per image. For 400 images per client, transmitting raw images requires $240.8$ MB, while transmitting the activations requires $61.44$ MB. The device-side LoRA adapter further incurs about $2.62$ MB, resulting in a total SFL communication overhead of approximately $64.1$ MB when both activation transmission and model updates are considered.}
    \footnotetext[2]{The computational memory usage was measured using the tools provided by FVCore.}

\section{Architecture and Workflow}\label{section2}
    \subsection{System Overview} We consider a collaborative split fine-tuning framework for LAMs over wireless edge networks. As shown in Fig.~\ref{fig:framework}, the system consists of a central server and a set of distributed devices $\mathcal{V}=\{1,2,\dots,V\}$. Each device $n\in\mathcal{V}$ maintains a private labeled dataset $\mathcal{D}_n=\{(x_{n,i},y_{n,i})\}_{i=1}^{D_n}$, where $x_{n,i}$ and $y_{n,i}$ denote the input sample and its label. Training proceeds over $T$ communication rounds, where a subset of active devices $\mathcal{N}\subseteq\mathcal{V}$ participates in each round. To accommodate resource-constrained devices, a pre-trained LAM with $E$ Transformer blocks is partitioned at a cut layer $e\in\{1,\dots,E\}$. The embedding layer and the first $e$ blocks are deployed on the device side, while the remaining $E-e$ blocks and the output head reside on the server. During training, the device performs a partial forward pass and uploads intermediate activations to the server. The server completes the remaining forward and backward computation and returns gradients to update the device-side parameters.

    Let $\boldsymbol{\omega}=\{\boldsymbol{\omega}^{\mathcal{C}},\boldsymbol{\omega}^{\mathcal{S}}\}$ denote the frozen backbone parameters on the device and server sides, respectively. Each participating device maintains LoRA adapters $\Delta\boldsymbol{\omega}_{n}=\{\Delta\boldsymbol{\omega}_{n}^{\mathcal{C}},\Delta\boldsymbol{\omega}_{n}^{\mathcal{S}}\}$, while the backbone remains fixed. For supervised fine-tuning, the global objective is
    \begin{equation}
        \min_{\Delta\boldsymbol{\omega}} F(\Delta\boldsymbol{\omega})
        = \sum_{n=1}^{\mathcal{N}} \rho_n F_n(\Delta\boldsymbol{\omega}),
    \end{equation}
    where $\rho_n = \frac{D_n}{\sum_{n \in \mathcal{M}_t} D_n}$ is the normalized data weight and $F_n(\Delta\boldsymbol{\omega})$ is the empirical loss on $\mathcal{D}_n$. The key challenge is that the optimization variables are lightweight, but the exchanged intermediate activations are still large. TSFLora addresses this mismatch by compressing the token representation at the split layer before transmission.

    \subsection{TSFLora Procedure}
        To support efficient fine-tuning of large models under resource constraints, TSFLora splits the model between the device and the server, applies token compression to reduce communication overhead, and adopts LoRA to lower memory usage. 

        \subsubsection{Model Initialization and Partitioning} At the beginning of training, the server broadcasts the device-side submodel $\boldsymbol{\omega}^{\mathcal{C}}$ and the corresponding LoRA initialization $\Delta\boldsymbol{\omega}_{n}^{\mathcal{C}}$ to the selected devices. The device stores the embedding layer and the first $e$ transformer blocks, while the server stores the remaining model. LoRA adapters are inserted into both submodels, yielding $\Delta \boldsymbol{\omega}_n^{\mathcal{C}} = \{(\mathbf{U}_1,\mathbf{V}_1),\dots,(\mathbf{U}_e,\mathbf{V}_e)\}$ and $\Delta \boldsymbol{\omega}_n^{\mathcal{S}} = \{(\mathbf{U}_{e+1},\mathbf{V}_{e+1}),\dots,(\mathbf{U}_E,\mathbf{V}_E)\}$.

        \subsubsection{Forward Propagation} In round $t$, device $n$ computes local activations on its dataset $\mathcal{D}_n$ using the device-side model with LoRA fine-tuning, given by
            \begin{equation}
                \mathbf{A}_n^t = F_n^{\mathcal{C}}\left(\boldsymbol{\omega}^{\mathcal{C}} + \Delta \boldsymbol{\omega}_n^{\mathcal{C}},\ \mathcal{D}_n\right).
            \end{equation}
        Here, $\mathbf{A}_n^t \in \mathbb{R}^{B \times (M+1) \times D}$, where $B$ is the batch size, $M+1$ consists of $M$ patch tokens and one \texttt{[CLS]} token, and $D$ is the embedding dimension. TSFLora then compresses $\mathbf{A}_n^t$ using the method in Section~\ref{section3}. Specifically, the token sequence is reduced to $K+2$ tokens by selection and merging, and the resulting activations are quantized from 32 bits to $q$ bits. The compressed activations are denoted by $\tilde{\mathbf{A}}_n^t \in \mathbb{R}^{B \times (K+2) \times D}$. After receiving $\tilde{\mathbf{A}}_n^t$, the server completes the remaining forward pass: $\hat{\mathbf{Y}}_n^t = F_n^{\mathcal{S}}\left(\boldsymbol{\omega}^{\mathcal{S}} + \Delta \boldsymbol{\omega}_n^{\mathcal{S}},\ \tilde{\mathbf{A}}_n^t\right)$.
        This step reduces transmission cost and server-side token processing by sending only $K+2$ quantized tokens instead of the full activation sequence.

        \subsubsection{Backward Propagation} The server computes gradients based on $\hat{\mathbf{Y}}_n^t$ and the labels, and updates its LoRA adapters by
            \begin{equation}
                \Delta \boldsymbol{\omega}_{n,t}^{\mathcal{S},i}
                \leftarrow
                \Delta \boldsymbol{\omega}_{n,t}^{\mathcal{S},i-1}
                - \eta \mathbf{g}_{n,t}^{\mathcal{S},i}, \quad \forall n \in \mathcal{N},
            \end{equation}
        where $\eta$ is the learning rate and $\Delta \boldsymbol{\omega}_{n,t}^{\mathcal{S},0}=\Delta \boldsymbol{\omega}_{n-1,t}^{\mathcal{S},I}$. The server then sends the gradient with respect to $\tilde{\mathbf{A}}_n^t$ back to the device, which performs local backpropagation and updates its LoRA adapters as
            \begin{equation}
                \Delta \boldsymbol{\omega}_{n,t}^{\mathcal{C},i}
                \leftarrow
                \Delta \boldsymbol{\omega}_{n,t}^{\mathcal{C},i-1}
                - \eta \mathbf{g}_{n,t}^{\mathcal{C},i}, \quad \forall n \in \mathcal{N},
            \end{equation}
        where $\Delta \boldsymbol{\omega}_{n,t}^{\mathcal{C},0}=\Delta \boldsymbol{\omega}_{n,t}^{\mathcal{C}}$.
    \subsubsection{LoRA Aggregation} After $I$ local updates, participating devices upload their device-side LoRA adapters to the server for aggregation. The server applies FedAvg as $\Delta \boldsymbol{\omega}_{t}^{\mathcal{C}} = \sum_{n \in \mathcal{M}_t} \rho_n\, \Delta \boldsymbol{\omega}_{n,t}^{\mathcal{C},I}$. The aggregated adapters are redistributed in the next round, and the above process repeats until convergence.
   
\section{Token Compression Scheme}\label{section3}
    This section presents the two-stage token compression design in TSFLora. It first performs token selection and merging, and then applies low-bit quantization to the refined activations. This design reduces communication overhead and server-side processing cost.

    \subsection{Token-level Selection and Merging}\label{token-selection-strategy} We consider a ViT-style token sequence in which the \texttt{[CLS]} token aggregates global information. For device $n$ in round $t$, let $\mathbf{A}_{n,t}\!\in\!\mathbb{R}^{B\times(M+1)\times D}$ denote the activation tensor output by the last device-side transformer block, where $B$ is the batch size, $M$ is the number of patch tokens, and $D$ is the embedding dimension. For the $b$-th sample, the token sequence is $\mathbf{A}_{n,t}^{(b)}=[\mathbf{a}_{n,t,0}^{(b)},\mathbf{a}_{n,t,1}^{(b)},\ldots,\mathbf{a}_{n,t,M}^{(b)}]$, where $\mathbf{a}_{n,t,0}^{(b)}$ is the \texttt{[CLS]} token.
    \subsubsection{CLS-attention scoring} In the last device-side transformer block, the attention matrix is computed as $\mathbf{Att}_{n,t}^{(b)}=\mathrm{softmax}\!\left(\mathbf{Q}_{n,t}^{(b)}\mathbf{K}_{n,t}^{(b)\top}/\sqrt{D}\right)$, where $\mathbf{Q}_{n,t}^{(b)}$ and $\mathbf{K}_{n,t}^{(b)}$ denote the query and key matrices. The attention score from the \texttt{[CLS]} token to the $i$-th patch token is $\alpha_{n,t,i}^{(b)}=\exp(\mathbf{q}_{n,t,0}^{(b)}\!\cdot\!\mathbf{k}_{n,t,i}^{(b)})/\sum_{j=1}^{M}\exp(\mathbf{q}_{n,t,0}^{(b)}\!\cdot\!\mathbf{k}_{n,t,j}^{(b)})$,
    where $\mathbf{q}_{n,t,0}^{(b)}$ and $\mathbf{k}_{n,t,i}^{(b)}$ denote the query and key vectors of the \texttt{[CLS]} and the $i$-th patch token.

    \subsubsection{Top-$K$ token selection}
    Given a token budget $K_{n,t}$, we retain the \texttt{[CLS]} token and select the top-$K_{n,t}$ patch tokens according to their attention scores. For each sample, define the selected index set as $\mathcal{S}_{n,t}^{(b)}=\mathrm{Top}\text{-}K(\alpha_{n,t,1}^{(b)},\ldots,\alpha_{n,t,M}^{(b)})$. The selected token sequence is $\mathbf{A}_{n,t}^{\mathrm{sel}}=\big[\mathbf{a}_{n,t,0}^{(b)},\{\mathbf{a}_{n,t,i}^{(b)}\}_{i\in\mathcal{S}_{n,t}^{(b)}}\big]_{b=1}^{B}$, where $\mathbf{A}_{n,t}^{\mathrm{sel}} \in \mathbb{R}^{B\times(K_{n,t}+1)\times D}$.

    \subsubsection{Token merging} To preserve information from discarded tokens, we aggregate the remaining tokens into one merged token. Let $ \mathcal{I}_{n,t}^{(b)}=\{1,\ldots,M\}\setminus \mathcal{S}_{n,t}^{(b)}.$
    The merged token is computed by an attention-weighted average: $\mathbf{a}_{n,t,\mathrm{merge}}^{(b)}= \big(\sum_{i\in\mathcal{I}_{n,t}^{(b)}}\alpha_{n,t,i}^{(b)}\mathbf{a}_{n,t,i}^{(b)}\big)/\big(\sum_{i\in\mathcal{I}_{n,t}^{(b)}}\alpha_{n,t,i}^{(b)}\big).$ Finally, the refined token sequence is constructed as
        \begin{equation}
            \mathbf{A}_{n,t}^{\mathrm{ref}} = \Big[
            \mathbf{a}_{n,t,0}^{(b)},\{\mathbf{a}_{n,t,i}^{(b)}\}_{i\in\mathcal{S}_{n,t}^{(b)}},
            \mathbf{a}_{n,t,\mathrm{merge}}^{(b)}
            \Big]_{b=1}^{B},
            \label{eq:token_refined}
        \end{equation}
    where $\mathbf{A}_{n,t}^{\mathrm{ref}} \in \mathbb{R}^{B\times (K_{n,t}+2)\times D}$. To quantify the approximation introduced by selection and merging, the following lemma bounds the resulting activation distortion:
    \begin{myLem}[Selection-Induced Activation Distortion]\label{lemma:attn_ref}
        Denote $\Psi\triangleq\max_{b,i}\|\mathbf{A}_{n,t}[b,i,:]\|_2^2$.
        Under the merge-and-scatter refinement, it holds that
        $\left\|\mathbf{A}_{n,t}-\mathbf{A}^{\mathrm{ref}}_{n,t}\right\|_F^2\le 4\Psi\,(M-K_{n,t})\,B.$
    \end{myLem}
    
    \subsection{Bit-Level Token Quantization}\label{sec:bit_level}
    Token selection reduces the sequence length, but the refined tensor $\mathbf{A}_{n,t}^{\mathrm{ref}} \in \mathbb{R}^{B \times (K_{n,t}+2) \times D}$ can still be large because the embedding dimension $D$ remains unchanged. TSFLora therefore applies stochastic low-bit quantization before transmission. Let $\mathbf{A}_{n,t}^{\mathrm{ref}}$ denote the refined activation tensor. We quantize each scalar entry independently and use dynamic range normalization to adapt to activation magnitude changes across devices and rounds. Define $A_{n,t}^{\max} = \max_{b,k,d} \left| \mathbf{A}_{n,t}^{\mathrm{ref}}[b,k,d] \right|$ and $A_{n,t}^{\min} = \min_{b,k,d} \left| \mathbf{A}_{n,t}^{\mathrm{ref}}[b,k,d] \right|$. Given bit-width $q_{n,t}$, the interval $[A_{n,t}^{\min}, A_{n,t}^{\max}]$ is uniformly partitioned into $2^{q_{n,t}}-1$ levels $\{\chi_0, \chi_1, \dots, \chi_{2^{q_{n,t}}-1}\}$ with step size $\Delta_{n,t} = \frac{A_{n,t}^{\max} - A_{n,t}^{\min}}{2^{q_{n,t}} - 1}$. For any scalar entry $\mathbf{A}_{n,t}^{\mathrm{ref}}[b,k,d]$, suppose $ \chi_\phi \le \left| \mathbf{A}_{n,t}^{\mathrm{ref}}[b,k,d] \right| \le \chi_{\phi+1}$. The stochastic quantizer $\mathcal{Q}(\cdot;q_{n,t})$ is defined as
        \begin{equation}
            \mathcal{Q}
            \big(
            \mathbf{A}_{n,t}^{\mathrm{ref}}[b,k,d]
            \big)
            =
            \begin{cases}
            \operatorname{sign}
            \big(
            \mathbf{A}_{n,t}^{\mathrm{ref}}[b,k,d]
            \big)
            \chi_\phi,
            & \text{w.p. } \iota,\\[4pt]
            \operatorname{sign}
            \big(
            \mathbf{A}_{n,t}^{\mathrm{ref}}[b,k,d]
            \big)
            \chi_{\phi+1},
            & \text{w.p.} 1-\iota,
            \end{cases}
            \label{eq:stochastic_quant}
        \end{equation}
    where $\iota= \frac{\chi_{\phi+1}-\left|\mathbf{A}_{n,t}^{\mathrm{ref}}[b,k,d]\right|}{\chi_{\phi+1}-\chi_\phi}$. This stochastic quantizer is unbiased and satisfies the following variance bound:
    \begin{myLem}(Unbiased Quantization Scheme)
        \label{lemma-quant}
            A randomized mapping $\mathcal{Q}:\mathbb{R}^{d}\rightarrow\mathbb{R}^{d}$ is an unbiased quantization scheme if there exists $\delta$ such that
            $\mathbb{E}\!\left[\mathcal{Q}(\mathbf{A}_{n,t}^{\mathrm{ref}})\right]=\mathbf{A}_{n,t}^{\mathrm{ref}}$ and
            $\mathbb{E}\!\left[\left\|\mathcal{Q}(\mathbf{A}_{n,t}^{\mathrm{ref}})-\mathbf{A}_{n,t}^{\mathrm{ref}}\right\|_F^2\right]\le\delta\,\left\|\mathbf{A}_{n,t}^{\mathrm{ref}}\right\|_F^2$,
            where $d=B\cdot(K_{n,t}+2)\cdot D$ and $\delta=\frac{1+\sqrt{2d-1}}{2(2^{q_{n,t}}-1)}$.
    \end{myLem}
    \subsection{Overhead Analysis} This subsection analyzes how the design variables $e$, $K$, and $q$ affect system overhead, and explains why TSFLora can jointly improve communication and computational efficiency.
        \subsubsection{Communication overhead} For each sample, a device transmits $K+2$ tokens, where each token has embedding dimension $D$ and quantization bit-width $q$. With batch size $B$, the uplink payload is $B(K+2)Dq$ bits, excluding the small metadata required for quantization ranges and signs. Compared with full-precision transmission of all $M+1$ tokens, the compression ratio is approximately $\frac{q(K+2)}{32(M+1)}$. Hence, decreasing either $K$ or $q$ directly reduces uplink traffic. Since the gradient returned by the server has the same compressed dimensionality, the downlink overhead scales with $B(K+2)D$ as well.
        \subsubsection{Computation overhead} On the device side, each client processes $B$ samples with $M+1$ tokens through the first $e$ transformer blocks. With LoRA rank $r<D$, the per-round complexity scales as $\mathcal{O}(B(M+1)Dre)$, which is substantially lower than full fine-tuning. On the server side, only the compressed sequence is forwarded through the remaining $E-e$ blocks, yielding complexity $\mathcal{O}(B(K+2)Dr(E-e))$. Therefore, $e$ determines the device--server workload partition, while $K$ simultaneously affects communication volume and server-side token processing. This coupling is a key characteristic of TSFLora.

\section{Non-Convex Convergence Analysis} \label{section4}
    We analyze the convergence behavior of TSFLora under non-convex objectives. During split fine-tuning, token-level selection and stochastic quantization introduce additional compression-induced perturbations into gradient updates. Our goal is to characterize how these perturbations affect stationarity. Following~\cite{han2024convergence,li2019convergence,santurkar2018does}, we first state standard assumptions on the local non-convex objectives $F_n(\Delta \boldsymbol{\omega})$, where $\Delta\boldsymbol{\omega}$ denotes the trainable LoRA adapters and the pretrained backbone remains fixed.
        \begin{myAssu}[$S$-smoothness]\label{assump1}
            Each local objective $F_n$ is $S$-smooth. For all $\Delta\boldsymbol{\omega},\Delta \boldsymbol{v}\in\mathbb{R}^w$, we have
            $F_n(\Delta\boldsymbol{\omega})\le F_n(\Delta\boldsymbol{v})+\langle \nabla F_n(\Delta \boldsymbol{v}),\Delta\boldsymbol{\omega}-\Delta \boldsymbol{v}\rangle+\frac{S}{2}\|\Delta\boldsymbol{\omega}-\Delta \boldsymbol{v}\|_2^2.$
        \end{myAssu}
        \begin{myAssu}[Unbiased and Bounded Stochastic Gradients]\label{assump2}
            The stochastic gradient $\boldsymbol{g}_n(\cdot)$ of $F_n(\cdot)$ is unbiased and has bounded variance:
            $\mathbb{E}_{\xi_n \sim \mathcal{D}_n}\!\left[\boldsymbol{g}_n(\Delta\boldsymbol{\omega},\xi_n)\right]=\nabla F_n(\Delta\boldsymbol{\omega})$, and
            $\mathbb{E}_{\xi_n \sim \mathcal{D}_n}\!\left[\left\|\boldsymbol{g}_n(\Delta\boldsymbol{\omega},\xi_n)-\nabla F_n(\Delta\boldsymbol{\omega})\right\|_2^2\right]\le\sigma_n^2$.
        \end{myAssu}
        \begin{myAssu}[Data Heterogeneity]\label{assump4}
            The divergence between local and global gradients is bounded by $\epsilon^2$, i.e.,
            $\mathbb{E}\!\left[\|\nabla F_n(\Delta\boldsymbol{\omega})-\nabla F(\Delta\boldsymbol{\omega})\|_2^2\right]\le\epsilon^2$.
        \end{myAssu}
        \begin{myAssu}[Gradient Lipschitzness w.r.t. Activations]\label{assump5}
            There exists a constant $\gamma>0$ such that the stochastic gradient is $\gamma$-Lipschitz continuous with respect to the activation tensor $\mathbf{A}$, i.e.,
            $\left\|\boldsymbol{g}_n(\Delta\boldsymbol{\omega},\xi_n;\mathbf{A})-\boldsymbol{g}_n(\Delta\boldsymbol{\omega},\xi_n;\mathbf{A}')\right\|_2\le\gamma\left\|\mathbf{A}-\mathbf{A}'\right\|_2$.
        \end{myAssu}

        \begin{myLem}[Gradient Perturbation Induced by Activation Distortion]\label{lemma3}
            Under Assumptions~\ref{assump2} and~\ref{assump5}, for the chain $\mathbf{A}_{n,t}\!\to\!\mathbf{A}^{\mathrm{ref}}_{n,t}\!\to\!\tilde{\mathbf{A}}_{n,t}$, let $\tilde{\boldsymbol{g}}_{n,t}\triangleq\boldsymbol{g}_n(\Delta\boldsymbol{\omega},\xi_n;\tilde{\mathbf{A}}_{n,t})$. Then $\mathbb{E}\!\left[\left\|\tilde{\boldsymbol{g}}_{n,t}-\nabla F_n(\Delta\boldsymbol{\omega})\right\|_2^2\right]\le 2\sigma_n^2 + 2\gamma^2(1+\kappa)\delta\,\Lambda + 8\gamma^2\!\left(1+\frac{1}{\kappa}\right)\Psi(M-K_n)B$, where $\Psi\triangleq\max_{b,i}\|\mathbf{A}_{n,t}[b,i,:]\|_2^2$, $\delta=\frac{1+\sqrt{2d-1}}{2(2^q-1)}$, $\Lambda=\mathbb{E}\!\left[\|\mathbf{A}^{\mathrm{ref}}\|_F^2\right]$, and $\kappa>0$.
        \end{myLem}
        \begin{myTheo}\label{theorem:nonconvex}
            Under Assumptions~\ref{assump1}--\ref{assump5} and Lemma~\ref{lemma3} (following G.3 of~\cite{han2024convergence}), let $\Delta\boldsymbol{\omega}_t$ denote the aggregated device-side LoRA parameters at round $t$. For a constant stepsize $\eta\le\frac{1}{4S}$,
            \begin{equation}
                \frac{1}{T} \sum_{t=0}^{T-1} \eta \mathbb{E}\!\left[\left\|\nabla F(\Delta\boldsymbol{\omega}_t)\right\|^2\right] \leq \frac{4\left(F(\Delta\boldsymbol{\omega}_0)-F^*\right)}{TI} + \mathcal{R}(q, K).
            \end{equation}
            where $F^*$ is the minimum global objective value and
            \begin{align}
                \mathcal{R}(q,K)
                =&\; \frac{8VSI}{T}
                \sum_{t=0}^{T-1}\eta^2
                \sum_{n=1}^{V}
                \frac{\rho_n^2+1}{\upsilon_n}
                \Bigg[
                2\sigma_n^2
                + \underbrace{2\gamma^2(1+\kappa)\Lambda\delta(q)}_{\text{quantization error}} \notag\\
                &\quad + \underbrace{8\gamma^2\!\left(1+\frac{1}{\kappa}\right)\Psi B (M-K)}_{\text{token selection error}}
                \Bigg].
            \end{align}
            Here, $\Psi\triangleq\max_{b,i}\|\mathbf{A}_{n,t}[b,i,:]\|_2^2$, $\delta=\frac{1+\sqrt{2d-1}}{2(2^{q_{n,t}}-1)}$, and $\Lambda=\mathbb{E}[\|\mathbf{A}^{\mathrm{ref}}\|_F^2]$. Also, $\kappa>0$ is the Young-inequality parameter, $\upsilon_n$ is the participation level of client $n$, and $\rho_n=\frac{D_n}{\sum_{n\in\mathcal{M}_t}D_n}$ is the data weight.
        \end{myTheo}

\section{Problem Formulation and Design Insights} \label{section5}
    TSFLora is governed by three coupled variables: the split layer $e$, the transmitted token budget $K$, and the quantization bit-width $q$. The split layer $e$ determines the device--server workload partitioning, while $K$ and $q$ jointly control the communication payload and optimization fidelity. We first characterize the system constraints. For device $n$, a split layer $e$ is feasible if $e \in \mathcal{E}_n \triangleq \{e \in \{1,\dots,E\} \mid M(e) \le \Omega_n\}$, where $M(e)$ denotes the peak device-side memory up to layer $e$ and $\Omega_n$ is the available memory budget.\par
    Meanwhile, the communication budget is constrained by the uplink capacity. For a mini-batch of size $B$, the compressed activation payload is given by
        \begin{equation}
            C(K,q) = B (K+2) D q \quad \text{bits},
        \end{equation}
    which must satisfy $C(K,q) \le C_{\max}$ under a given uplink communication burden $C_{\max}$. Theorem~\ref{theorem:nonconvex} shows that the optimization error contains a compression-dependent term $\mathcal{R}(q,K)$. Increasing $K$ or $q$ generally improves optimization fidelity but incurs higher communication cost and server-side computation. This reveals an intrinsic trade-off between communication efficiency and model performance. Based on this observation, we introduce the following formulation to characterize the optimal operating point:
        \begin{align}
            \mathcal{P}: \quad
            &\min_{K,q} \quad \mathcal{R}(q,K) \\
            \text{s.t.} \quad
            & C(K,q) \le C_{\max}, \\
            & M(e) \le \Omega_n, \\
            & 1 \le K \le M, \\
            & q \in \mathcal{Q}.
        \end{align}
    This formulation is not intended to derive a closed-form or online optimization algorithm. Instead, it serves as an analytical framework to understand how the key design variables $(e,K,q)$ jointly affect system feasibility and learning performance. In particular, it highlights that $K$ and $q$ act as the primary communication control variables in split fine-tuning.\par
    Guided by this formulation, we explore a range of $(e,K,q)$ configurations in Section~\ref{section6} to identify practical operating points that balance memory constraints, communication cost, and accuracy. This design-oriented approach provides actionable insights for deploying communication-efficient split fine-tuning in resource-constrained edge systems.
        
        \begin{table}[t]
            \centering
            \caption{Resource Configuration of Virtual devices}
            \label{tab:device-config}
            \begin{tabular}{c|c|c}
            \hline
            \textbf{device ID} & \textbf{Compute Fraction} & \textbf{Memory Fraction} \\
            \hline
            0-2   & 0.05 & 0.08 \\
            3--6   & 0.10 & 0.10 \\
            7--9  & 0.15 & 0.12 \\
            \hline
            \end{tabular}
        \end{table}
        
        \begin{table}[t]
            \caption{Top-1 Accuracy (\%) after 50 communication rounds (best/second/third are bold/underlined/italic).}
            \centering
            \scriptsize
            \setlength{\tabcolsep}{2.5pt}
            \renewcommand{\arraystretch}{0.92}
            \begin{tabular}{c|c|cc|cc|cc}
            \hline
            \textbf{Backbone} & \textbf{Method} 
            & \multicolumn{2}{c|}{CIFAR-10}
            & \multicolumn{2}{c|}{CIFAR-100}
            & \multicolumn{2}{c}{TinyImageNet} \\
            \cline{3-8}
             & & IID & Non & IID & Non & IID & Non \\
            \hline
            \multirow{8}{*}{ViT-Small}
            & LocalLoRA & 91.11 & 89.06 & 53.97 & 43.54 & 38.48 & 26.42 \\
            & FedLoRA & 95.02 & 90.41 & 69.53 & 54.01 & 46.14 & 33.11 \\
            & SplitLoRA & 95.63 & 94.72 & 82.23 & 82.00 & 70.98 & 68.73 \\
            & SFLora & \textit{97.06} & \underline{96.47} & \textit{86.01} & \textbf{85.55} & \textbf{77.30} & \textbf{76.83} \\
            & SFLora (8-bit) & \textbf{97.35} & \textbf{96.65} & \underline{86.04} & \underline{85.54} & \underline{76.88} & \underline{75.90} \\
            & SFLora (4-bit) & 96.08 & 94.98 & 81.55 & 79.97 & 73.12 & 71.41 \\
            \rowcolor{cyan!10}
            & TSFLora (8-bit, 40 tokens) & \underline{97.13} & \textit{96.22} & \textbf{86.07} & \textit{85.11} & \textit{76.60} & \textit{75.80}\\
            \rowcolor{cyan!5}
            & TSFLora (8-bit, 30 tokens) & 95.90 & 95.90 & 84.99 & 83.96 & 75.02 & 74.17\\
            \hline
            \multirow{8}{*}{ViT-Base}
            & LocalLoRA & 94.91 & 91.96 & 58.32 & 46.05 & 44.92 & 31.32 \\
            & FedLoRA & 96.99 & 93.96 & 79.10 & 64.42 & 61.64 & 42.60 \\
            & SplitLoRA & \textbf{98.10} & \textit{97.60} & 89.58 & \textit{89.49} & 83.23 & 83.00 \\
            & SFLora & \textit{97.94} & \underline{97.76} & 88.07 & \underline{89.52} & \underline{84.52} & \underline{84.35} \\
            & SFLora (8-bit) & \underline{98.08} & \textbf{97.91} & \textbf{90.32} & \textbf{89.97} & \textbf{85.00} & \textbf{84.67} \\
            & SFLora (4-bit) & 97.62 & 97.17 & \underline{89.93} & 87.10 & 82.96 & 82.35 \\
            \rowcolor{cyan!10}
            & TSFLora (8-bit, 40 tokens) & 97.87 & 97.51 & \textit{89.77} & 89.04 & \textit{83.93} & \textit{83.60}\\
            \rowcolor{cyan!5}
            & TSFLora (8-bit, 30 tokens) & 97.88 & 97.22 & 88.97 & 88.30 & 82.85 & 82.84\\
            \hline
            \multirow{8}{*}{ViT-Large}
            & LocalLoRA & 96.42 & 94.92 & 74.59 & 63.77 & 60.64 & 38.76 \\
            & FedLoRA & \textbf{98.16} & 97.57 & 87.33 & 77.91 & 75.24 & 43.12 \\
            & SplitLoRA & \textit{98.11} & \underline{97.77} & \textbf{90.60} & \underline{90.09} & \textbf{87.17} & \underline{86.95} \\
            & SFLora & 97.91 & \textit{97.69} & 89.87 & \textit{89.52} & 86.34 & 85.74 \\
            & SFLora (8-bit) & \underline{98.15} & \textbf{97.79} & \underline{90.58} & \textbf{90.11} & \underline{86.85} & \textbf{87.05} \\
            & SFLora (4-bit) & 97.97 & 97.59 & \textit{89.89} & 89.30 & \textit{86.43} & \textit{86.35} \\
            \rowcolor{cyan!10}
            & TSFLora (8-bit, 40 tokens) & 97.98 & 97.47 & 89.76 & 89.35 & 85.60 & 85.28\\
            \rowcolor{cyan!5}
            & TSFLora (8-bit, 30 tokens) & 97.87 & 97.14 & 89.07 & 88.23 & 84.45 & 83.83\\
            \hline
            \end{tabular}
            \label{tab:big_comparison}
        \end{table}

        \begin{figure}[t]
                \centering
                \includegraphics[width=0.9\linewidth]{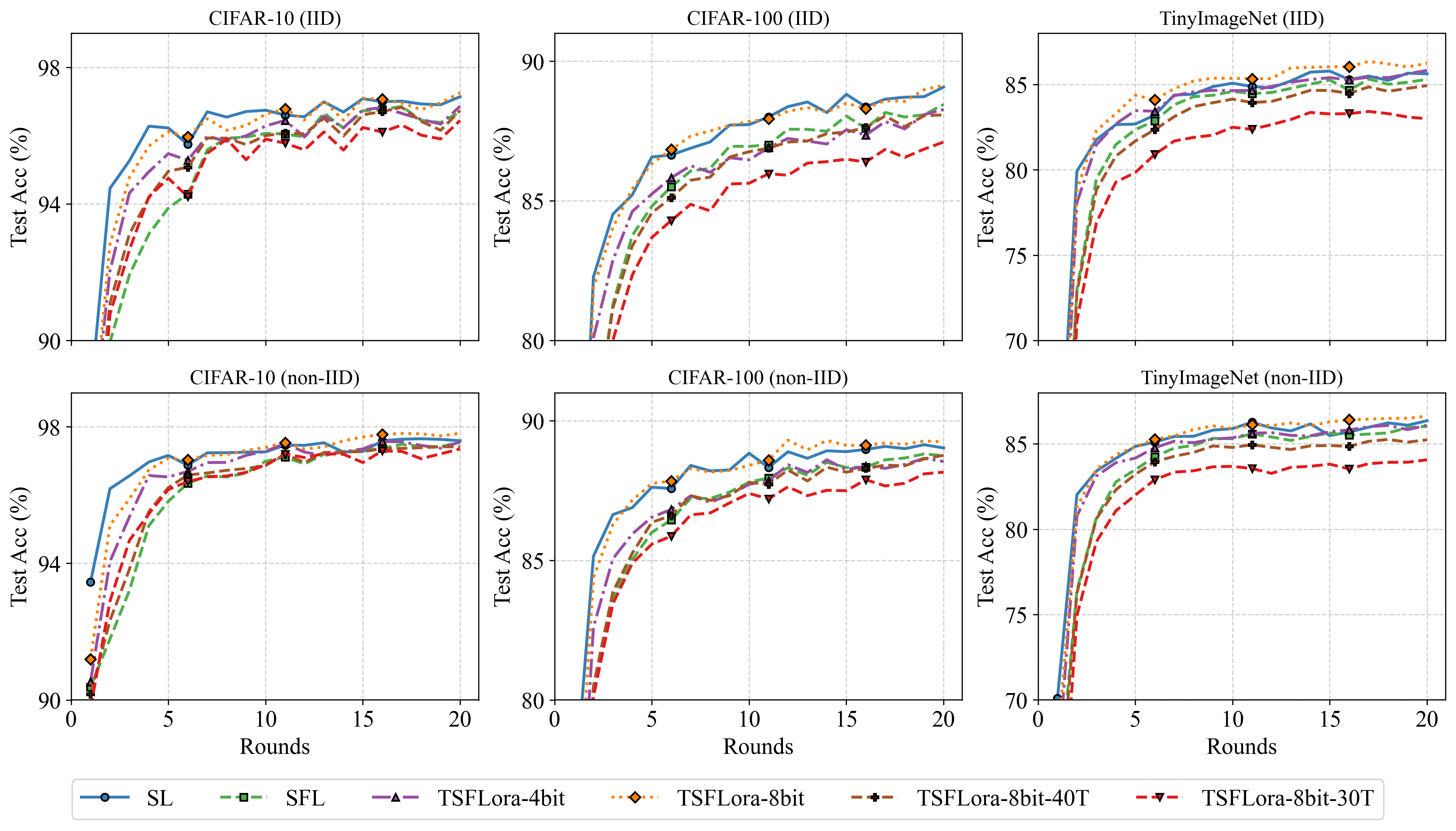}
                \caption{Testing accuracy with random selection of 10 out of 50 devices under different Dirichlet distributions.}
                \label{fig:convergence}
            \end{figure}
        
        \begin{figure}[t]
          \centering
          \subfloat[\tiny CIFAR-100 ($e$=2)]{\includegraphics[width=0.3\linewidth]{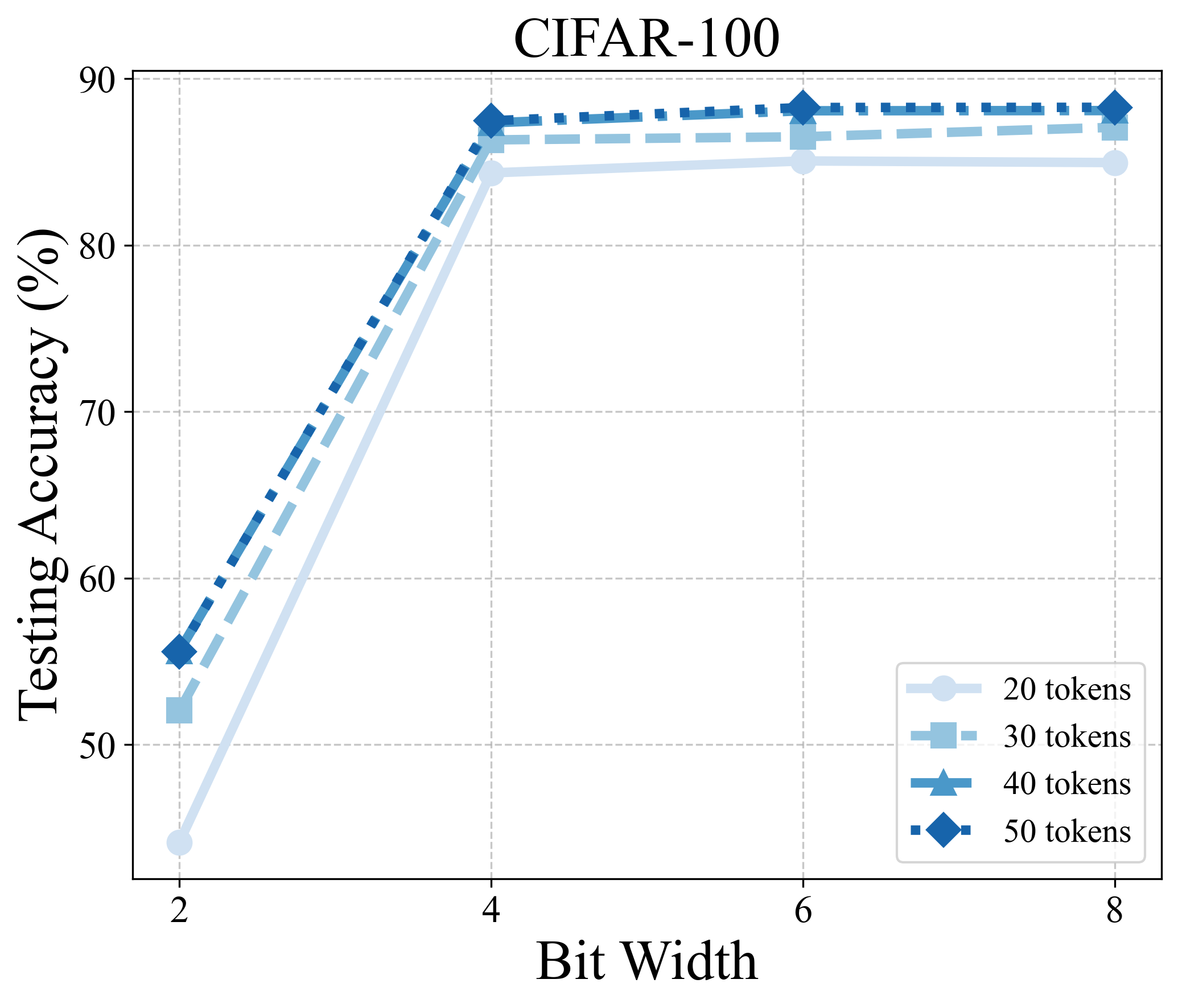}\label{fig:cifar100-token}}
          \subfloat[\tiny CIFAR-100 (50 Tokens)]{\includegraphics[width=0.3\linewidth]{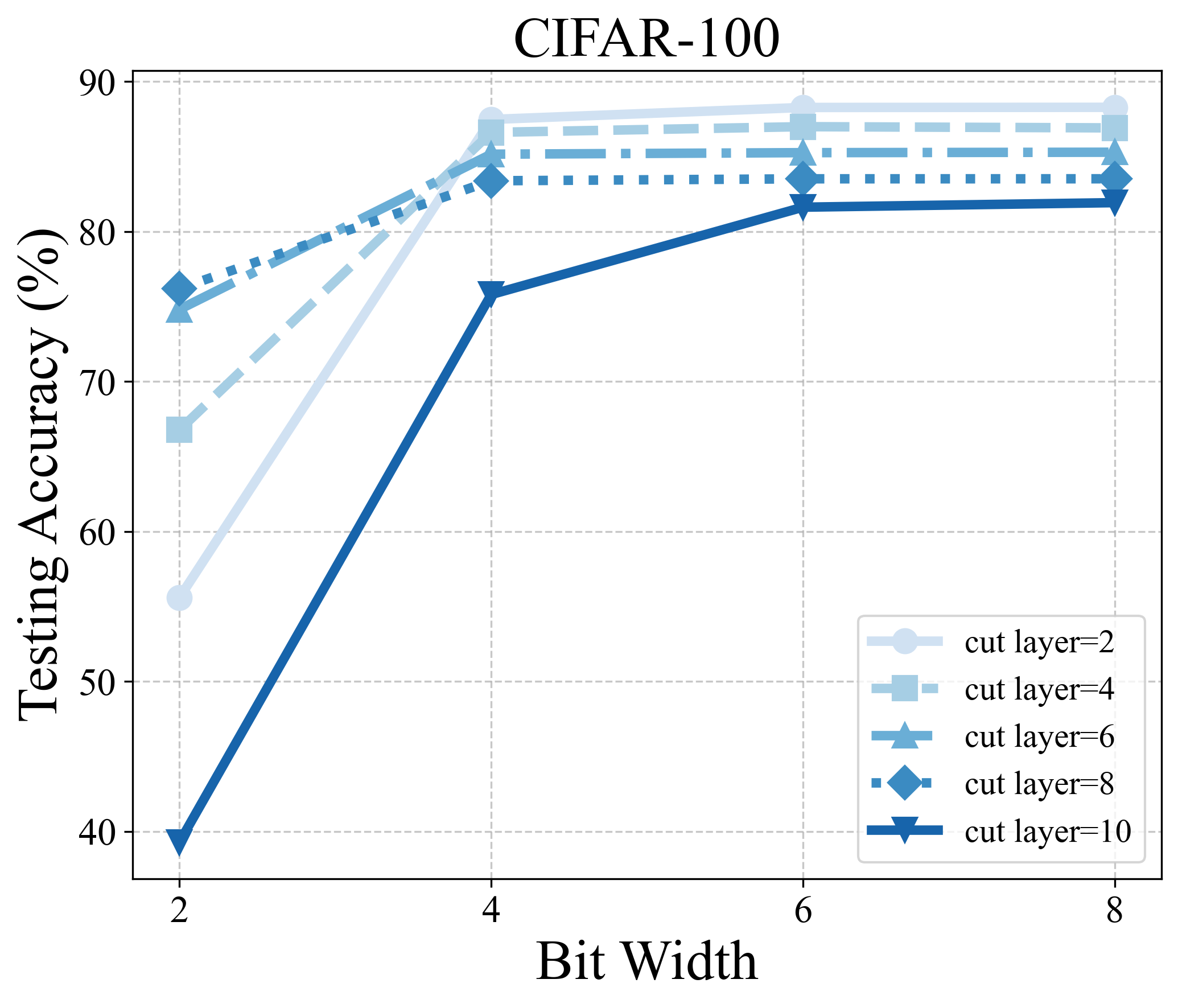}\label{fig:cifar100-cut}} 
          \subfloat[\tiny Patch 16$\times$16]{\includegraphics[width=0.3\linewidth]{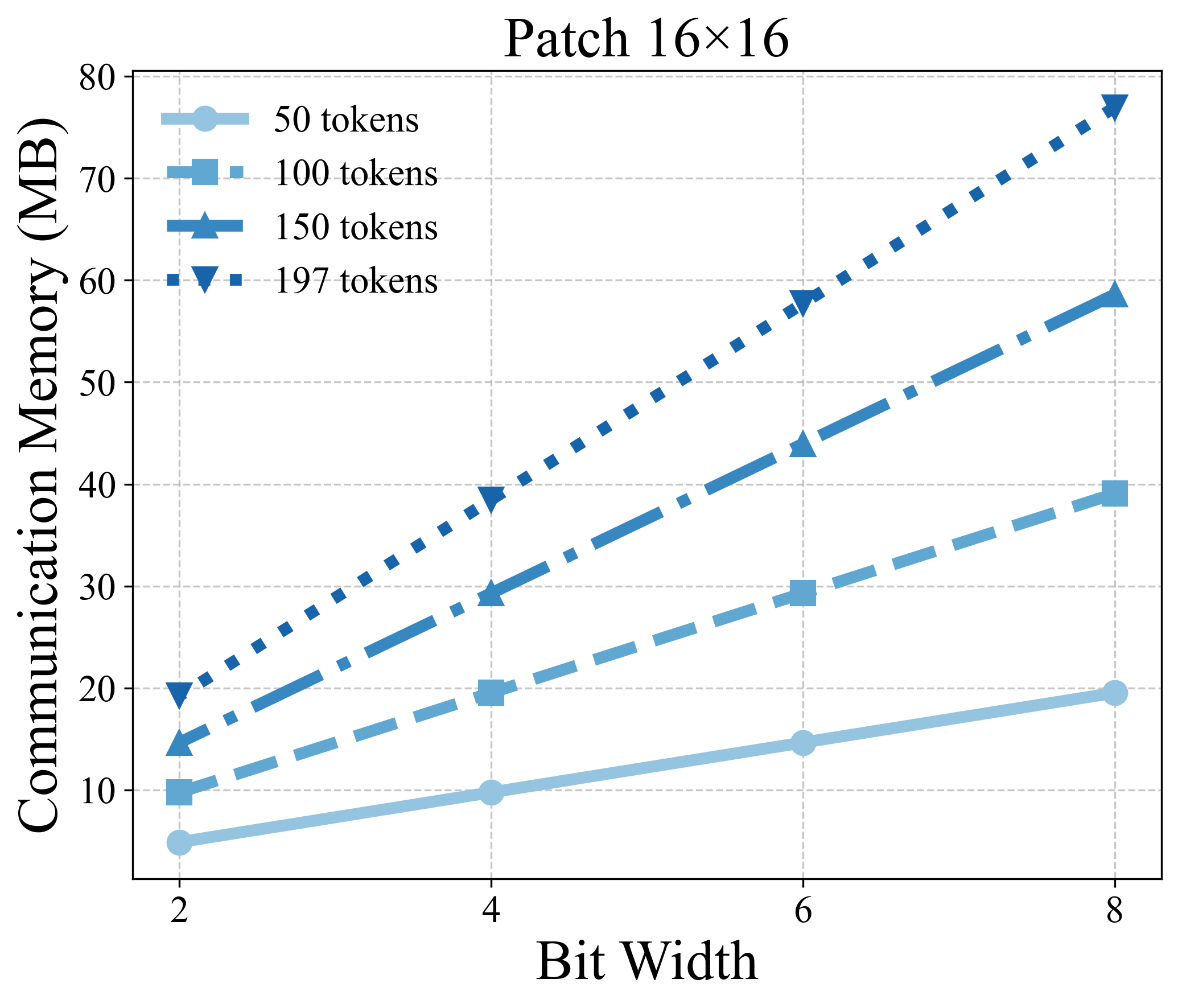}\label{fig:patch16}}
          \hfill
          \subfloat[\tiny TinyImageNet ($e$=2)]{\includegraphics[width=0.3\linewidth]{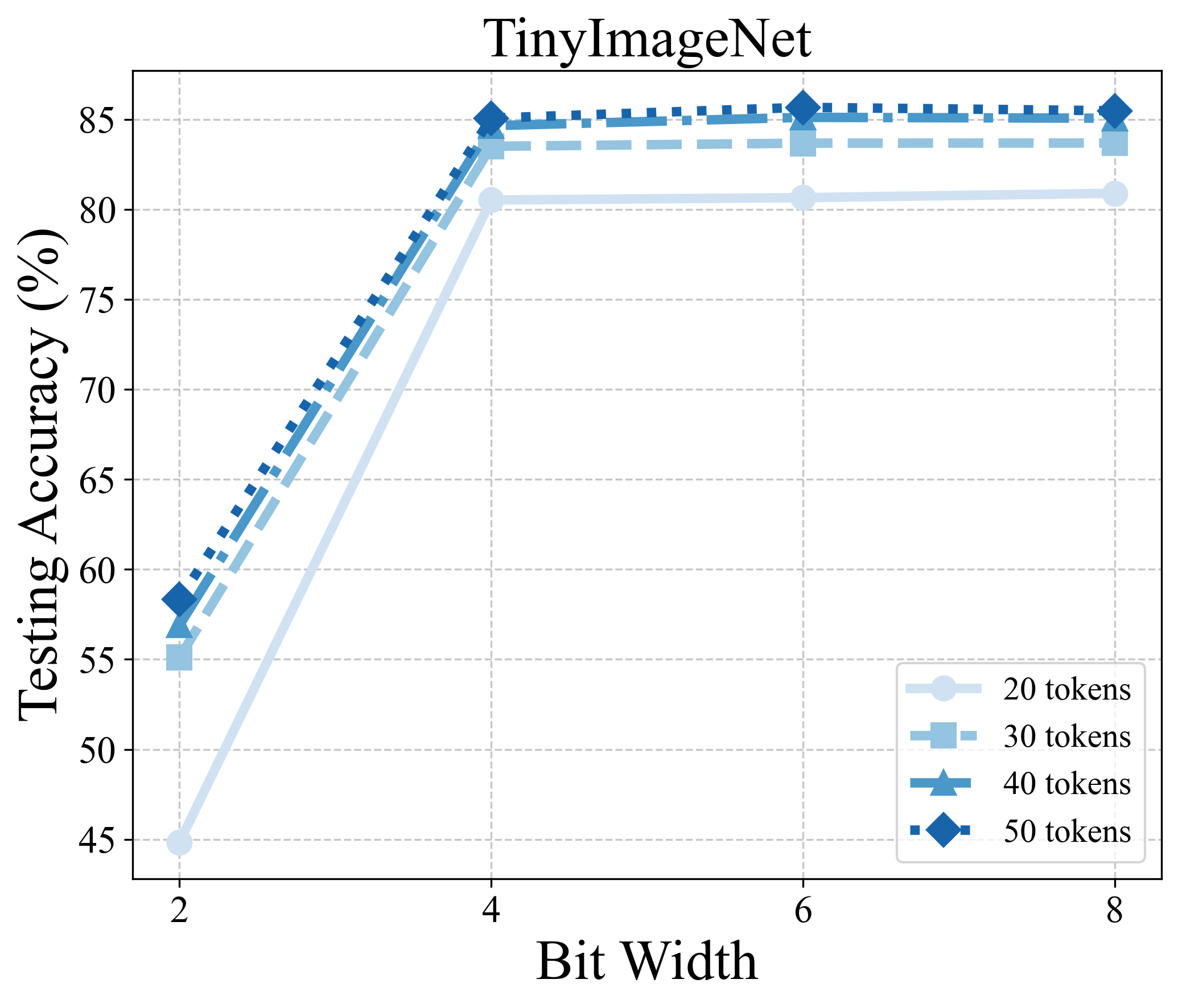}\label{fig:tiny-token}} 
          \subfloat[\tiny TinyImageNet (50 Tokens)]{\includegraphics[width=0.3\linewidth]{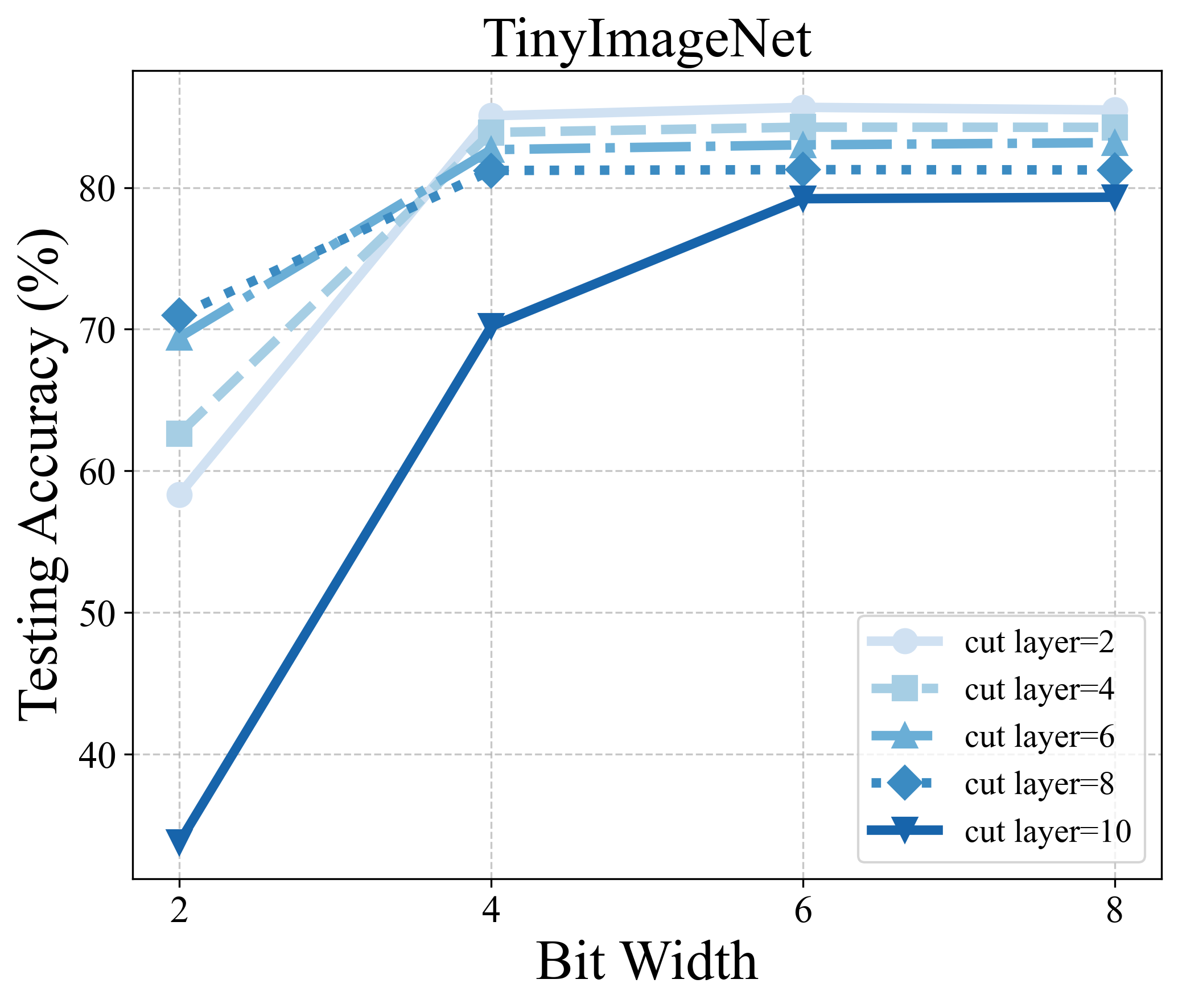}\label{fig:tiny-cut}}
          \subfloat[\tiny Patch 32$\times$32]{\includegraphics[width=0.3\linewidth]{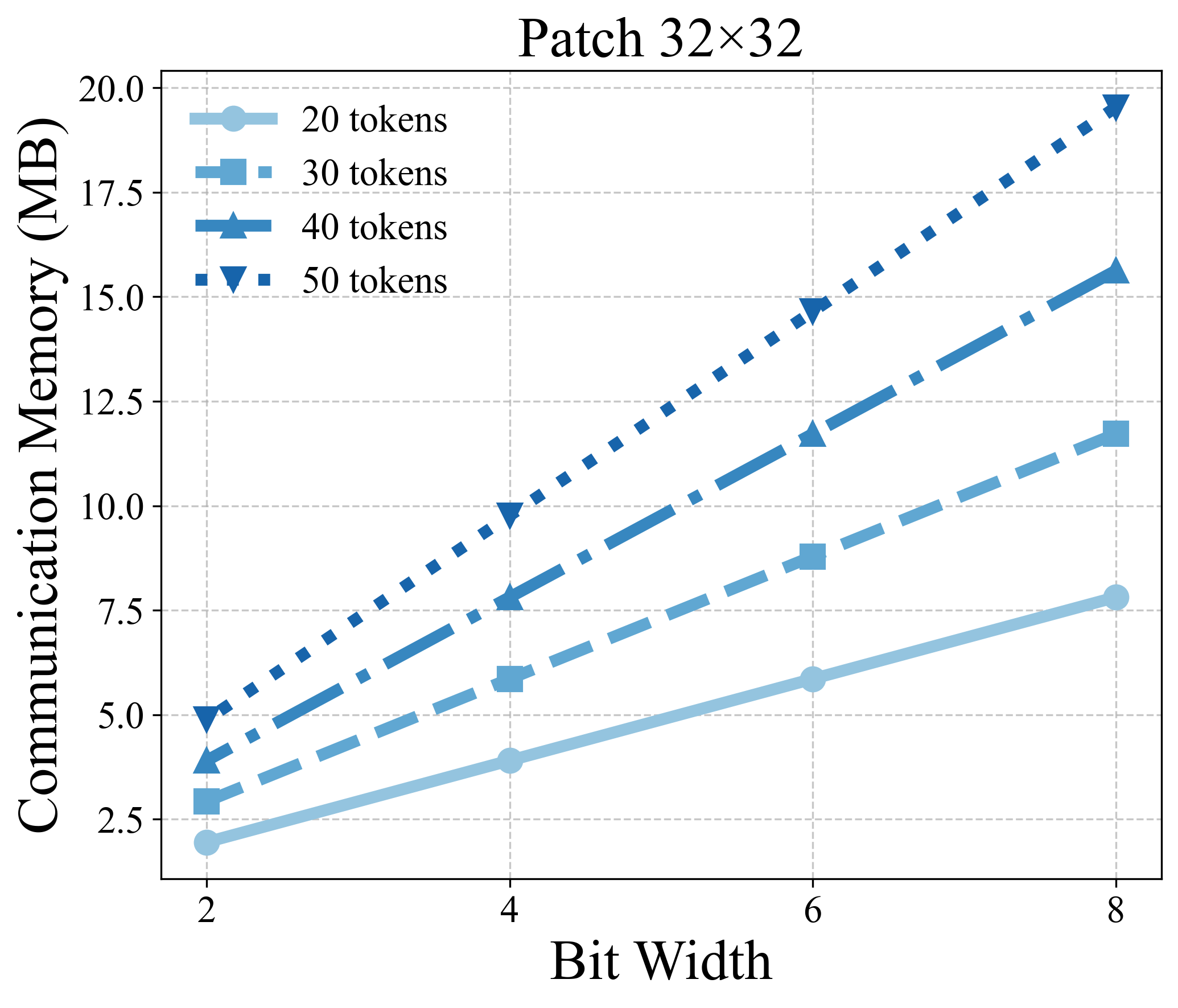}\label{fig:patch32}}
          \caption{Accuracy and communication memory across token numbers, bit widths, and cut layer selections on CIFAR-100 and TinyImageNet.}
          \label{fig:bitwidth_token_cut}
        \end{figure}

        \begin{figure}[t]
          \centering
          \subfloat[\tiny Device Peak Memory (MB)]{
            \includegraphics[width=0.30\linewidth]{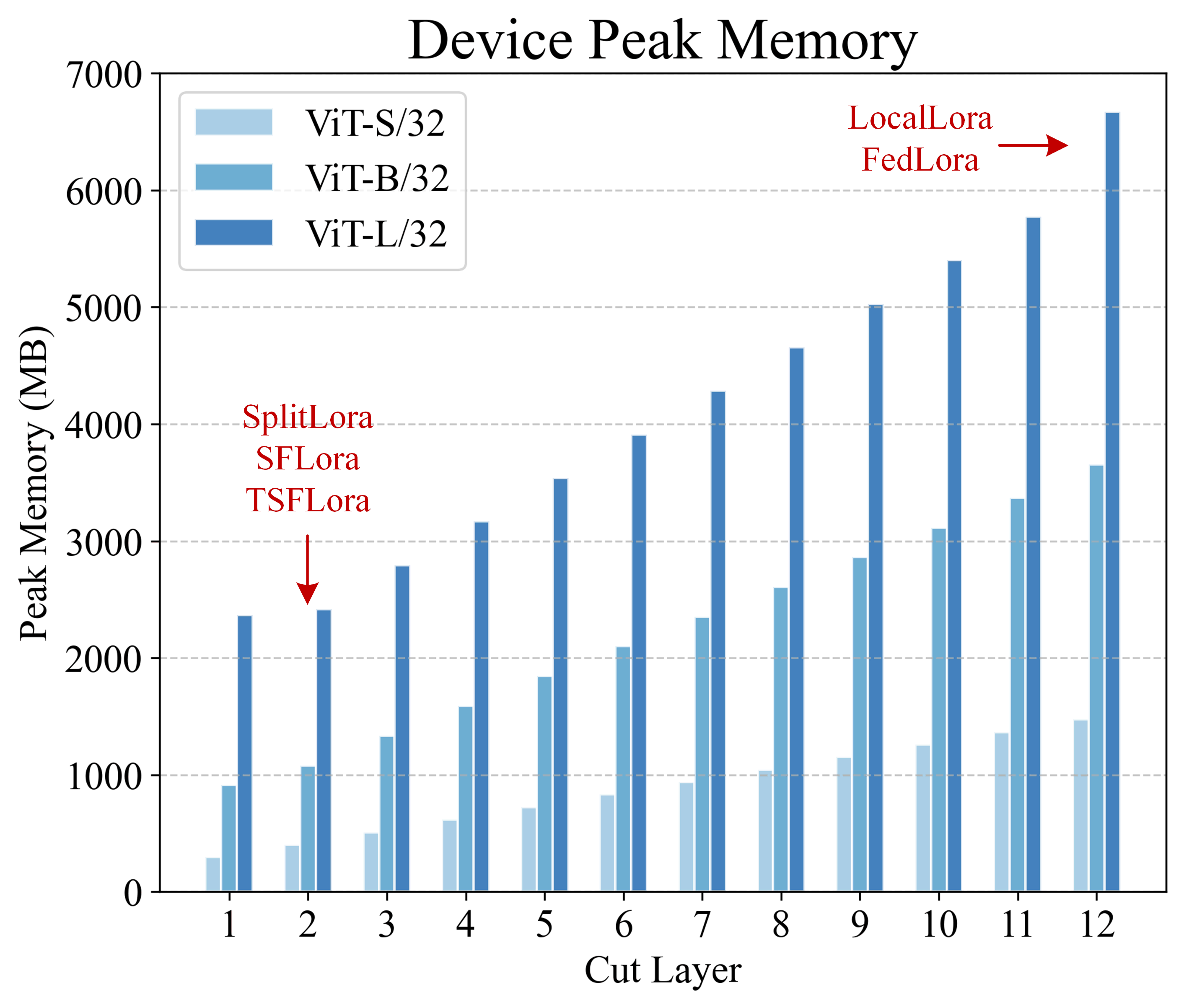}
            \label{5a}
          }
          \subfloat[\tiny Communication Overhead]{
            \includegraphics[width=0.30\linewidth]{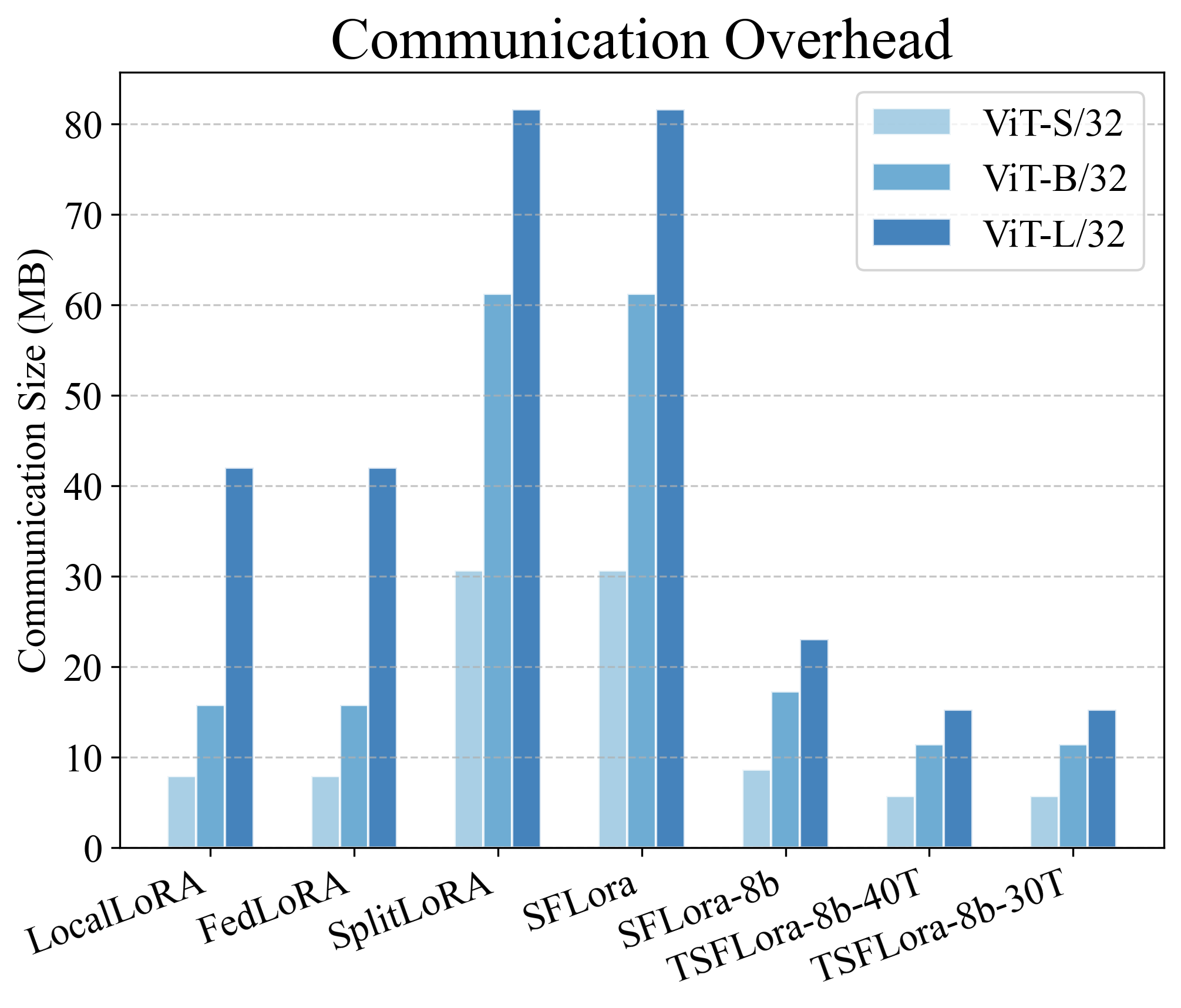}
            \label{5b}
          }
        \subfloat[\tiny Execution Time Comparison @ 10 Mbps]{
            \includegraphics[width=0.30\linewidth]{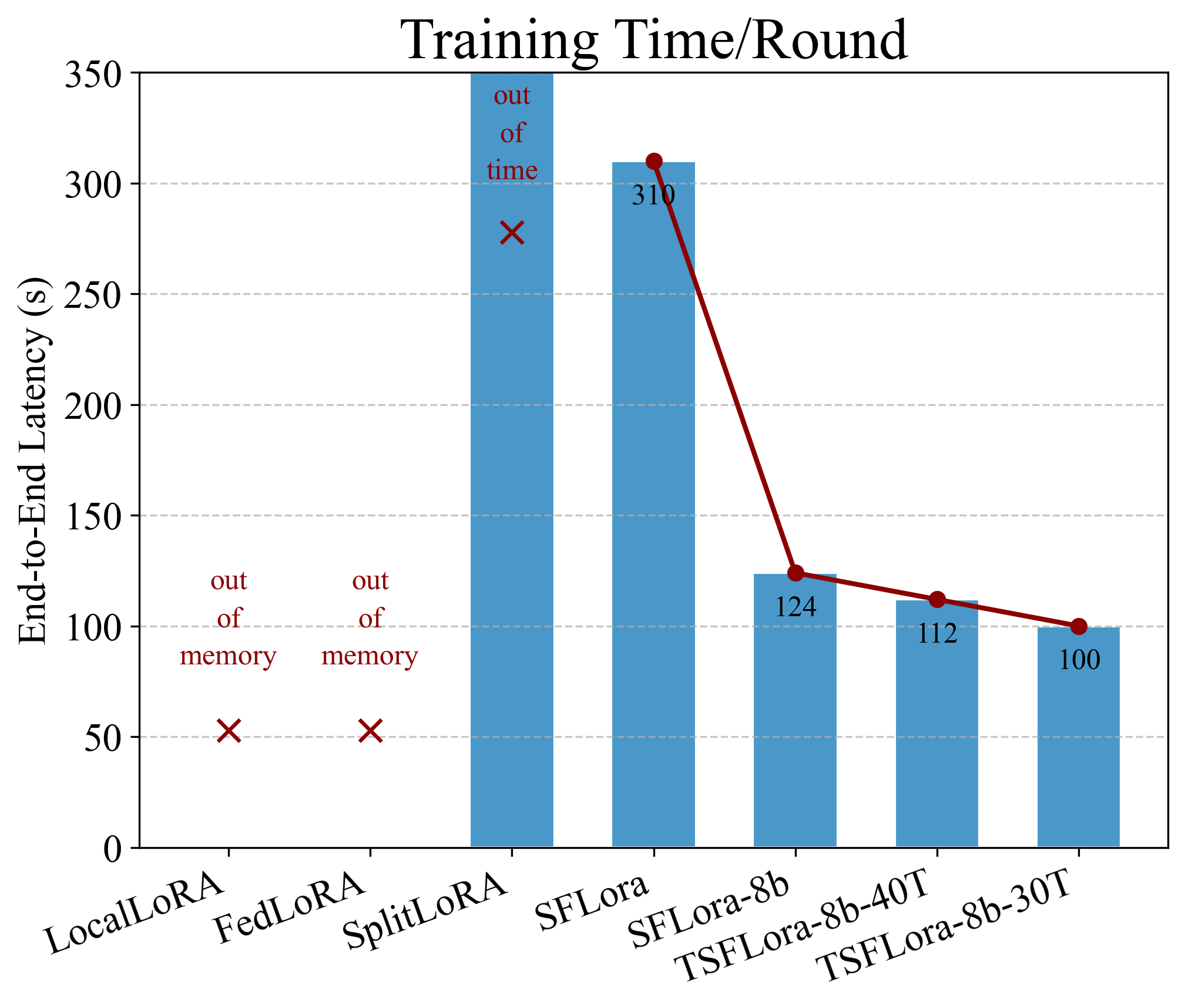}
            \label{5c}
          }
          \hfill
           \subfloat[\tiny Execution Time under @ 40 Tokens]{
            \includegraphics[width=0.30\linewidth]{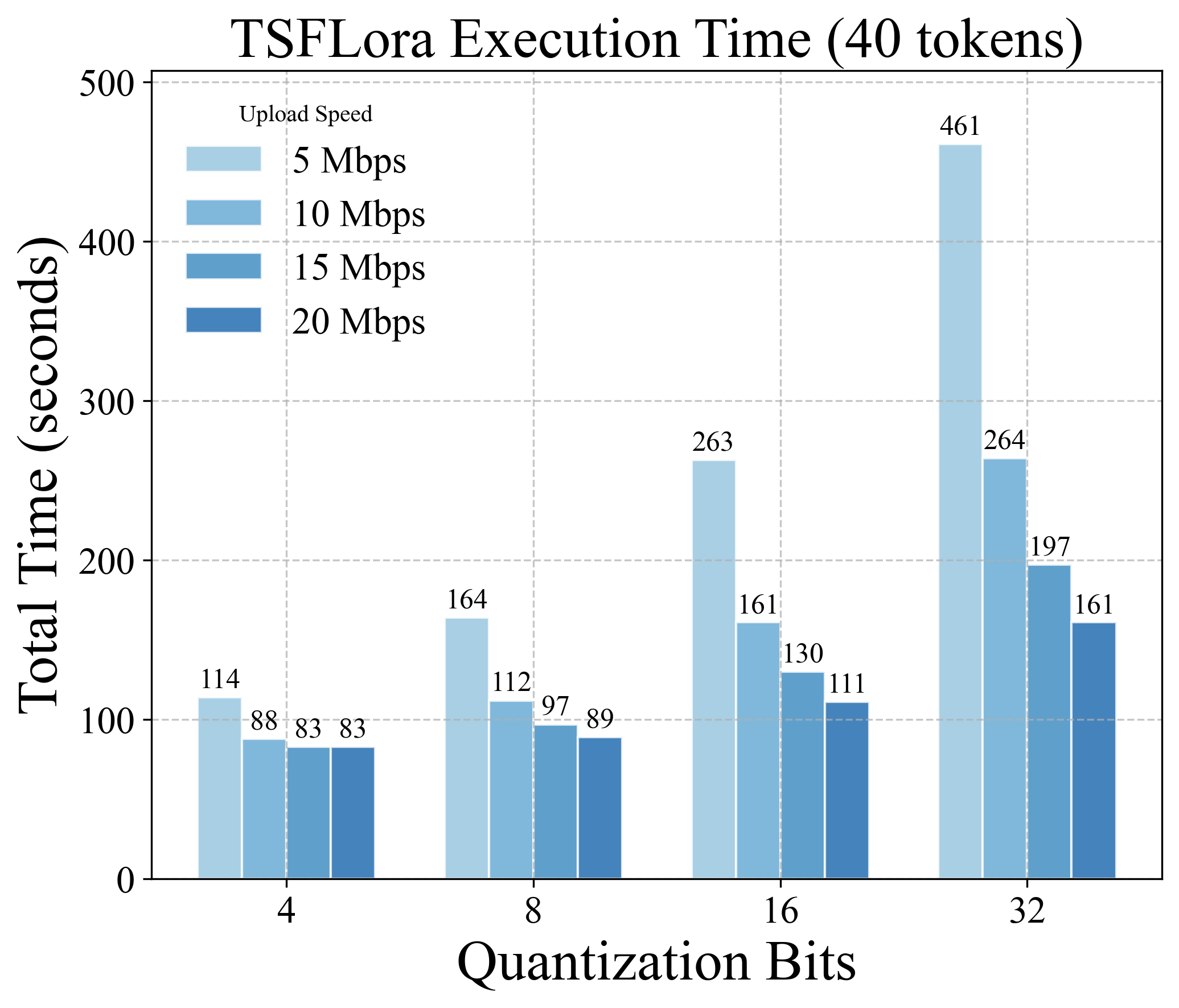}
            \label{5d}
          }
          \subfloat[\tiny Execution Time Heatmap @ 10 Mbps]{
            \includegraphics[width=0.30\linewidth]{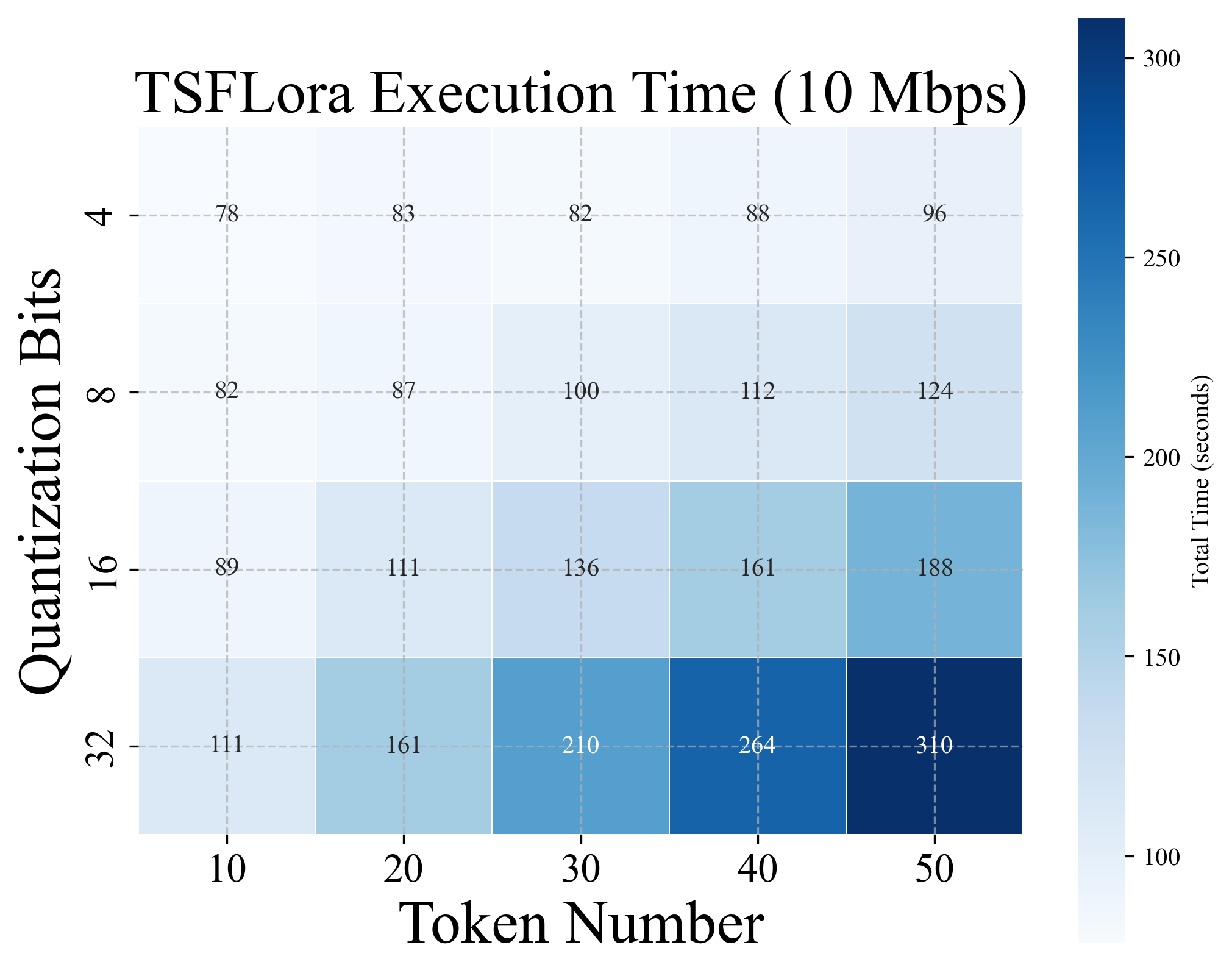}
            \label{5e}
          }
          \subfloat[\tiny Execution Time Heatmap @ 15 Mbps]{
            \includegraphics[width=0.30\linewidth]{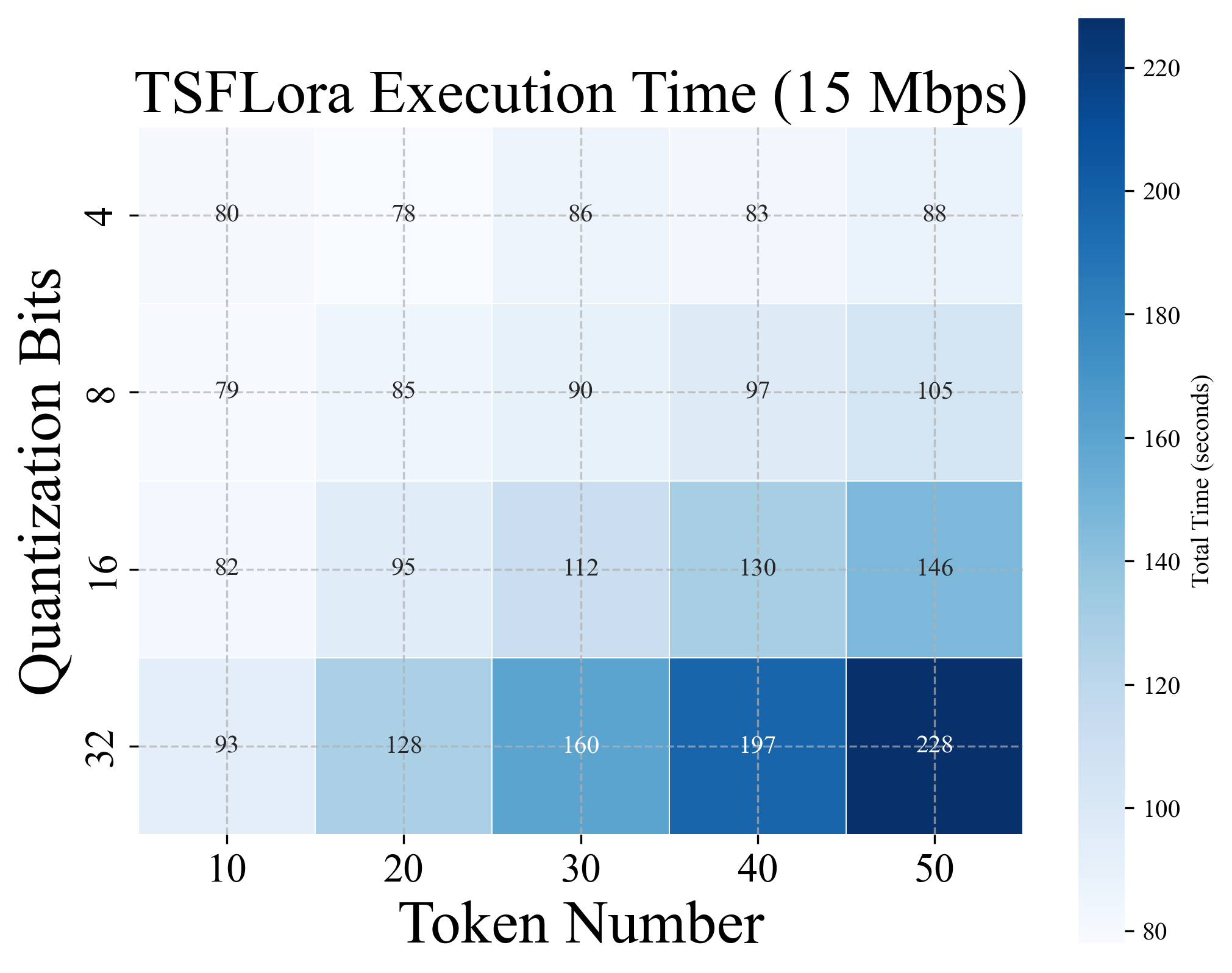}
            \label{5f}
          }
          \caption{System profiling results: device peak memory, communication cost, and total execution time under different quantization and token selection settings.}
          \label{fig:system-overhead}
        \end{figure}

\section{Experiments}\label{section6}
        \subsection{System Implementation} 
            We implement \textbf{TSFLora} on a distributed testbed with one server and multiple heterogeneous devices using two NVIDIA L20 GPUs (48\,GB each). One GPU acts as the server, while the other is virtualized into ten device instances to emulate resource-constrained edge devices with different capabilities (Table~\ref{tab:device-config}). Devices communicate with the server over Wi-Fi via WebSocket.\par
            We evaluate TSFLora on distributed image classification with \textit{ViT-Small/32}, \textit{ViT-Base/32}, and \textit{ViT-Large/32}, using pre-trained weights from \textit{timm}. The learning rate, batch size, and LoRA rank are set to $0.1$, $64$, and $32$, respectively. Experiments are conducted on CIFAR-10, CIFAR-100, and TinyImageNet under both IID and non-IID settings, where the latter is generated by a Dirichlet distribution with $\alpha=0.5$. We compare TSFLora with \textit{LocalLoRA}, \textit{FedLoRA}, \textit{SplitLoRA}, \textit{SFLora}, and two quantized variants, \textit{SFLora (8-bit)} and \textit{SFLora (4-bit)}. TSFLora further integrates token selection and low-bit quantization to reduce both communication and server-side computation overhead.

        \subsection{Performance Evaluation} 
            Table~\ref{tab:big_comparison} reports the top-1 accuracy after 50 communication rounds. The results show three consistent trends. First, split fine-tuning methods outperform conventional federated baselines because they keep the adaptation process closer to centralized fine-tuning. Second, moderate activation quantization introduces only limited accuracy degradation: \textit{SFLora (8-bit)} remains close to, even better than full-precision \textit{SFLora} across all backbones and datasets. Third, TSFLora preserves most of this accuracy while further reducing the token payload. With 40 transmitted tokens, TSFLora achieves accuracy comparable to \textit{SFLora (8-bit)}. Reducing the token number to 30 leads to a slight accuracy degradation, but yields further reduction in communication overhead.\par
            Figure~\ref{fig:convergence} shows the convergence behavior. TSFLora converges slightly slowly than \textit{SFLora (8-bit)} because of stronger compression, but it still reaches competitive final accuracy. For example, on CIFAR-100 with ViT-Base/32 under non-IID data, \textit{TSFLora (8-bit, 40 tokens)} achieves $89.04\%$ accuracy, close to \textit{SFLora (8-bit)} at $89.97\%$, while using a smaller transmitted representation.\par

        \subsection{Model Effectiveness}
            We next study how $K$, $q$, and $e$ affect the accuracy-efficiency trade-off. Figures~\ref{fig:bitwidth_token_cut}\subref{fig:cifar100-token} and \ref{fig:bitwidth_token_cut}\subref{fig:tiny-token} indicate that accuracy improves from 2-bit to 8-bit but saturates beyond 4-bit, while reducing tokens from 50 to 40 or 30 causes only minor degradation.\par

            Figures~\ref{fig:bitwidth_token_cut}\subref{fig:cifar100-cut} and \ref{fig:bitwidth_token_cut}\subref{fig:tiny-cut} further illustrate the interaction between quantization and the split layer.  Under aggressive 2-bit quantization, shallower and deeper split layers suffer from more severe accuracy degradation, indicating higher sensitivity to quantization errors. In contrast, middle splits exhibit more stable performance in this regime. As the bit-width increases to 4 or 8 bits, the performance gap across different split layers becomes much smaller, suggesting that sufficient activation precision can mitigate the impact of split-layer selection. Overall, these results indicate that the optimal split depends jointly on quantization fidelity and device constraints, rather than following a fixed preference for shallow or deep partitions.\par

            Finally, Figures~\ref{fig:bitwidth_token_cut}\subref{fig:patch16} and \ref{fig:bitwidth_token_cut}\subref{fig:patch32} show that communication memory increases with both bit-width and token number. For example, under the $32\times32$ patch setting, reducing the token number from 50 to 30 decreases communication memory from about 20 MB to 12 MB, which is close to a $40\%$ reduction with little accuracy loss. These results confirm that token selection and quantization are complementary: one reduces the number of transmitted elements, and the other reduces the number of bits per element.

        \subsection{System Effectiveness}
            We now examine whether the proposed compression gains translate into system-level benefits. Figure~\ref{fig:system-overhead}\subref{5a} shows that TSFLora reduces peak device memory relative to the baselines and remains within a typical 4 GB edge-device budget even with 8-bit precision and 40 tokens. This confirms that model splitting and LoRA make large-model adaptation feasible on memory-constrained devices.\par
            Figure~\ref{fig:system-overhead}\subref{5b} shows that TSFLora achieves the lowest communication overhead across the evaluated model scales, particularly with 8-bit or 4-bit quantization and 30 or 40 tokens. Under 4-bit quantization with 30 tokens, the communication volume is reduced by more than $80\%$. Figure~\ref{fig:system-overhead}\subref{5c} further shows that the communication reduction directly translates into lower end-to-end training latency under a 10 Mbps uplink. In contrast, \textit{LocalLoRA} and \textit{FedLoRA} are infeasible because of memory constraints, while \textit{SplitLoRA} exceeds the latency budget.\par
            Figure~\ref{fig:system-overhead}\subref{5d} evaluates latency across uplink bandwidths from 5 Mbps to 20 Mbps. TSFLora remains practical even at low bandwidth, and its latency becomes less sensitive to bandwidth when 4-bit quantization is used because the payload is small. Figures~\ref{fig:system-overhead}\subref{5e} and \ref{fig:system-overhead}\subref{5f} also show the expected limit of aggressive compression: very small budgets, such as 2-bit quantization with 10 tokens, further reduce latency but noticeably hurt accuracy. Therefore, the experiments support the formulation in Section~\ref{section5}: useful operating points arise from balancing memory feasibility, communication budget, and optimization fidelity.

\section{Conclusion}\label{section7}
    This paper presented TSFLora, a token-compressed split fine-tuning framework for large-model adaptation in bandwidth- and memory-constrained edge systems. By combining split learning, LoRA, attention-guided token selection/merging, and low-bit activation quantization, TSFLora jointly reduces activation communication overhead and server-side processing cost. Experiments on CIFAR-10, CIFAR-100, and TinyImageNet under both IID and non-IID settings show up to $6.8\times$ communication reduction and $41\%$ memory savings with competitive accuracy, indicating that split-layer token compression is a practical direction for communication-efficient edge fine-tuning.

\bibliographystyle{IEEEtran}
\bibliography{main.bib}

\end{document}